\documentclass[aps,prl,twocolumn,superscriptaddress,a4paper,english,longbibliography]{revtex4-2}

\usepackage{times}
\usepackage{color}
\usepackage[colorlinks=true,urlcolor=blue,citecolor=blue]{hyperref}
\usepackage{url}
\usepackage{breakurl}
\usepackage{soul}
\usepackage{graphicx}
\usepackage{amsmath}
\usepackage{subfigure}
\usepackage{xcolor}
\usepackage{amssymb}
\usepackage{latexsym}
\usepackage{hyperref}
\DeclareGraphicsExtensions{.png,.eps}
\usepackage{float}
\usepackage{ulem}

\usepackage{xcolor}
\usepackage[urlcolor=blue]{hyperref}      
\hypersetup{
    colorlinks = true,                    
    citecolor = {blue},
    linkcolor = {purple},
           }
\begin{document}	
	
\title{Deep learning Local Reduced Density Matrices for Many-body Hamiltonian Estimation}

\author{Xinran Ma}
\affiliation{Department of Physics, Beijing Normal University, Beijing 100875, China}
\author{Z. C. Tu}
\affiliation{Department of Physics, Beijing Normal University, Beijing 100875, China}
\author{Shi-Ju Ran}\email[Corresponding author. Email: ] {sjran@cnu.edu.cn}
\affiliation{Department of Physics, Capital Normal University, Beijing 100048, China}

\date{\today}

\begin{abstract}
Human experts cannot efficiently access the physical information of quantum many-body states by simply ``reading'' the coefficients, but have to reply on the previous knowledge such as order parameters and quantum measurements. In this work, we demonstrate that convolutional neural network (CNN) can learn from the coefficients of local reduced density matrices to estimate the physical parameters of the many-body Hamiltonians, such as coupling strengths and magnetic fields, provided the states as the ground states. We propose QubismNet that consists of two main parts: the Qubism map that visualizes the ground states (or the purified reduced density matrices) as images, and a CNN that maps the images to the target physical parameters. By assuming certain constraints on the training set for the sake of balance, QubismNet exhibits impressive powers of learning and generalization on several quantum spin models. While the training samples are restricted to the states from certain ranges of the parameters, QubismNet can accurately estimate the parameters of the states beyond such training regions. For instance, our results show that QubismNet can estimate the magnetic fields near the critical point by learning from the states away from the critical vicinity. Our work illuminates a data-driven way to infer the Hamiltonians that give the designed ground states, and therefore would benefit the existing and future generalizations of quantum technologies such as Hamiltonian-based quantum simulations and state tomography.
\end{abstract}

\maketitle

\section{Introduction}

Machine learning (ML) has recently been applied to various issues that are difficult using purely the ``conventional'' techniques in physics (for instance, tensor network \cite{VMC08MPSPEPSRev, O19TNrev, ran2020tensor}, quantum Monte Carlo \cite{CA86QMCrev, nightingale1998quantum}, and etc.). The successful applications include identifying the classical/quantum phases and topologies without computing order parameters  \cite{wang2016discovering, carrasquilla2017machine, van2017learning, zhang2018machine, rem2019identifying, rodriguez2019identifying, scheurer2020unsupervised}, predicting physical properties of materials \cite{rupp2012fast, xie2018crystal, hanakata2018accelerated}, efficiently representing non-trivial quantum states \cite{carleo2017solving, choo2018symmetries, glasser2018neural, deng2017machine}, to name but a few.

Among others, obtaining the eigenstates, particularly the ground states, of a given quantum many-body Hamiltonian belongs to the central topics in the contemporary physics \cite{avella2012strongly, kuramoto2020quantum}. The inverse problems, which are of equal significance and practicality, are much less studied due to the lack of valid methods. ML serves as a novel approach that has recently gained certain inspiring successes in such problems \cite{fournier2020artificial, teoh2020machine, hanakata2020forward, arsenault2017projected}. In particular, one important issue under hot debate is to access the information of the potentials or interactions by learning from physical data. For instance, Xin \textit{et al} utilized fully-connected neural network to recover the ground states of $k$-local Hamiltonians from local measurements \cite{xin2019local}. Hegde \textit{et al} employed the kernel ridge regression to achieve accurate and transferable predictions of Hamiltonians for a variety of material environments \cite{hegde2017machine}. Li \textit{et al} identified the effective Hamiltonians in magnetic systems and extracted the dominant spin interactions in MnO and TbMnO$_3$ through multiple linear regression \cite{li2020constructing}. Sehanobish \textit{et al} proposed the quantum potential neural networks to reconstruct the effective potential given the wave functions \cite{sehanobish2020learning}.

However, most existing works in this direction utilize the regression methods or shallow neural networks, which usually possess relative low learning or generalizing powers. In ML, one usually uses deep networks, such as convolutional neural network (CNN) \cite{lecun1989backpropagation, krizhevsky2017imagenet}, to solve sophisticated problems such as the classifications of real-life images. The excellent learning and generalization abilities of CNN have been widely recognized in numerous applications in computer sciences (c.f. Refs. \cite{aloysius2017review, yao2019review, sultana2020review} for instance). Recently, Berthusen \textit{et al} utilized CNN to extract the crystal field Stevens parameters from the thermodynamic data, which illustrates the validity of CNN-based method in deducing physical information \cite{berthusen2020learning}. Goh \textit{et al} put forward a deep CNN model named Chemception to predict chemical properties with the 2D drawings of molecules \cite{goh2017chemception}. Laanait \textit{et al} used an encoder-decode architecture with convolutional layers to generate the local electron density of material by learning from the diffraction patterns \cite{laanait2019exascale}. It is interesting and unexplored whether CNN is capable of solving more challenging issues, including those with the presence of strong correlations and many-body effects.

\begin{figure}[tbp]
	\centering
	\includegraphics[angle=0,width=1\linewidth]{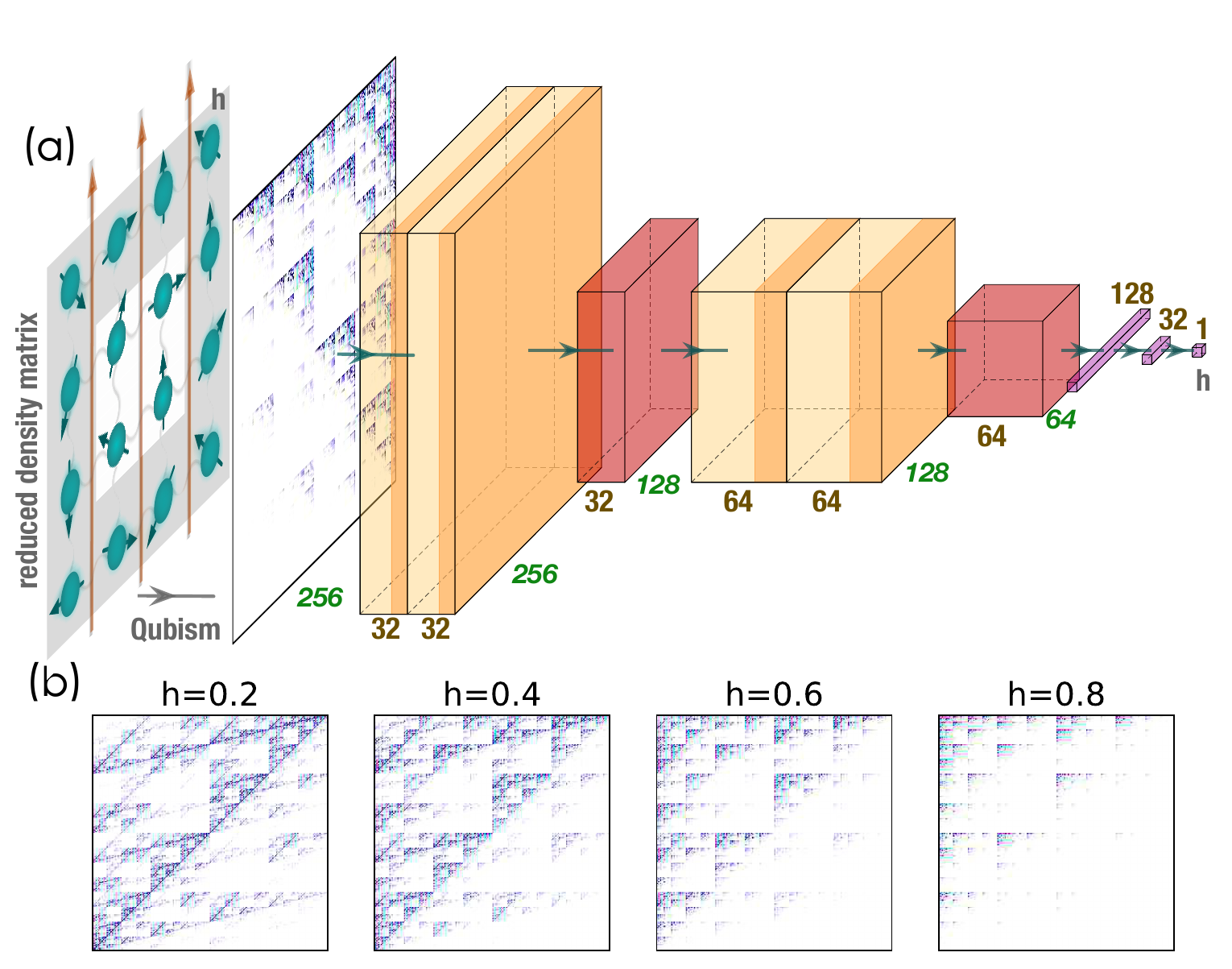}
	\caption{(Color online) (a) Illustration of QubismNet. Its first part is the Qubism map that transforms a quantum wave-function to an image of fractals. The second part is a convolutional neural network that maps the images to the estimations of the target parameters. (b) Examples of the images obtained by Qubism. The ground states are from the XY model with the magnetic fields $h=0.2 \sim 0.8$.}
	\label{fig-QN}
\end{figure}

In this work, the problem we consider is inverse to those of solving the eigenstates of a given Hamiltonian. Suppose the state or its local reduced density matrices (RDMs) is given. Our aim is to estimate the parameters in the many-body Hamiltonian by training a CNN model, so that the given state is the ground state. Solving such a problem would be meaningful and important for, e.g., designing the Hamiltonian of a quantum annealer to prepare a target state \cite{das2008colloquium}. To this end, we propose QubismNet that consists of two main parts (FIG. \ref{fig-QN} (a)). The first part is a map to transform the RDMs into images, and the second is a CNN to transform the images to the estimations of the target parameters in the Hamiltonian. The purpose of mapping the RDMs to images is to utilize the power of CNN processing images. Similar idea has been used in Ref. \cite{berthusen2020learning}, where the thermodynamic data (specific heat and others) are transformed to images by wavelet transformation before being fed to the CNN. The RDM-image map we use is known as Qubism \cite{rodriguez2012qubism}, where the obtained images are of fractals (see some examples in FIG. \ref{fig-QN} (b)) that can reveal the physical properties of the state.

We benchmark QubismNet on several quantum spin models defined on 1D and 2D lattices. The learning and generalization powers of QubismNet are tested by dividing the samples (i.e., the ground states taking different values of certain parameter in the Hamiltonian) into testing and generalizing sets. The parameters corresponding to the states in the testing set are independently and identically distributed (\textit{i.i.d.}) as the states in the training set. QubismNet can estimate the parameters of such states with high performance. The parameters corresponding to the states in the generalizing set are restricted in a certain range in which no training states are taken. To keep the training data balanced, the training samples are taken from the boundaries of the whole parameter space, and the generalizing samples are taken from a sub-region in the middle of the parameter space. Our results show that QubismNet can generalize what it has learned from the training set to estimate the parameters of the generalizing states with fair performance. For instance, QubismNet only learns from the states away from the critical point and well estimates the magnetic fields given the RDMs of the states in the critical vicinity. Our work suggests that CNN is capable of extracting information directly from the coefficients of the RDMs, while human experts have to reply on the previous knowledge such as order parameters and measurements.

\section{QubismNet: estimating physical parameters from local reduced density matrices}
\label{sec-QubismNet}

QubismNet consists of two main parts. The first part is a Qubism map \cite{rodriguez2012qubism} that was originally proposed to transforms states to images of fractals in a one-to-one way.

Taking the quantum Ising model (QIM) in a transverse magnetic field as an example, the Hamiltonian reads
\begin{equation}\label{H_Ising}
\hat{H}(h) = J \sum_{\left \langle i, j \right \rangle} \hat{S}_i^z \hat{S}_j^z - h \sum_{k=1}^{L} \hat{S}_k^x,
\end{equation}
with $\hat{S}^{\alpha}$ the $\alpha$-component spin operator ($\alpha = x, z$) and $L$ the system size. Here, we take the coupling constant $J=1$ as the energy scale. There are several well-established methods to calculate the ground states given the Hamiltonians, such as density matrix renormalization group (DMRG) \cite{white1992density, white1993density}, tensor network algorithms \cite{O19TNrev, VMC08MPSPEPSRev, ran2020tensor}, quantum Monte Carlo \cite{CA86QMCrev, nightingale1998quantum}, and etc. This work considers an inverse problem: estimating the magnetic fields $h$ (or other parameters) given the RDMs of the ground states.

In specific, we denote the training set as $\{|\psi_m \rangle\}$ ($m=1, \ldots, N_{\text{train}}$), where $|\psi_m \rangle$ is the ground state of $\hat{H}(h_m)$. To train the CNN whose output is the value of the target parameter, we choose the mean-square error (MSE) as the loss function
\begin{equation} \label{eq-loss}
\varepsilon = \frac{1}{N_{\text{train}}} \sum_{m=1}^{N_{\text{train}}} (h^p_m - h_m)^2,
\end{equation}
with $h^p_m$ the estimation of the magnetic field of $|\psi_m \rangle$ by the QubismNet. The variational parameters in the CNN are optimized by minimizing the loss function using the gradient method. We choose RMSProp \cite{hinton2012neural} as the optimizer to control the gradient steps. More details of the Qubism map and CNN are provided in the Supplementary Material.

To benchmark the generalization power of QubismNet, we introduce the testing and generalizing sets \footnote{Note that the datasets, i.e., the ground states of the Hamiltonians with different physical parameters, are prepared by the exact diagonalization algorithm for small sizes or by DMRG algorithm for large sizes}. The testing sets contains the states whose magnetic fields are different from but \textit{i.i.d.} with the training states. The states in the generalizing set are different from both the training and testing states, and are distributed in a different region. For instance, we uniformly choose $N_{\text{train}}$ values of $h$ within $0<h<0.5-\delta/2$ and $0.5+\delta/2<h<1$ for the training set, and choose other $N_{\text{test}}$ values of $h$ in the same regions of $h$ for the testing set. For the generalizing set, we uniformly choose $N_{\text{g}}$ values of $h$ within $0.5-\delta/2<h<0.5+\delta/2$. We dub $\delta$ as \textit{generalization width}. Note $h=0.5$ is the quantum critical point of QIM. The states in either the testing or the generalizing sets are not used to train the CNN.


For the large-size systems, it is inefficient to directly apply the Qubism map, as it requires the full coefficients of the quantum states. To avoid such a problem, we bring in the RDMs combined with purification. In specific, we choose a subsystem of a moderate size (denoted by $L_b$) and calculate the reduced density matrix $\hat{\rho}(|\psi_m\rangle) = \text{Tr}_s |\psi_m\rangle \langle \psi_m |$ with $\text{Tr}_s$ tracing the degrees of freedom in the subsystem. In general, if $L_b$ is comparable or larger than the correlation length, the RDM would contain the dominant physical information of the whole system \cite{verstraete2006matrix, zauner2015transfer}. Interestingly, according to our simulations, it is even not necessary to set $L_b$ larger than the correlation length to accurately estimate the physical parameters from the RDMs.

To map a RDM to image by Qubism, we write it as a pure state as $\hat{\rho}(|\psi_m\rangle) = \sum_{ii'} \rho_{ii'} |i\rangle \langle i'| \to |\rho_m\rangle= \sum_{ii'} \rho_{ii'} |i i'\rangle$. One can see that $|\rho_m\rangle$ is the purification of $\hat{\rho}(|\psi_m\rangle)^2$ since we have $\text{Tr}_{i'} |\rho_m\rangle\langle \rho_m| = \hat{\rho}(|\psi_m\rangle)^2$. Therefore, we feed the QubismNet by $\{|\rho_m\rangle\}$, which contain identical amount of information as $\{\hat{\rho}(|\psi_m\rangle)\}$. The parameter complexity of $|\rho_m\rangle$ is independent on the size of the whole system. It is same as the complexity of a state with $2L_b$ spins.

\begin{figure*}[tbp]
	\centering
	\includegraphics[angle=0,width=0.95\linewidth]{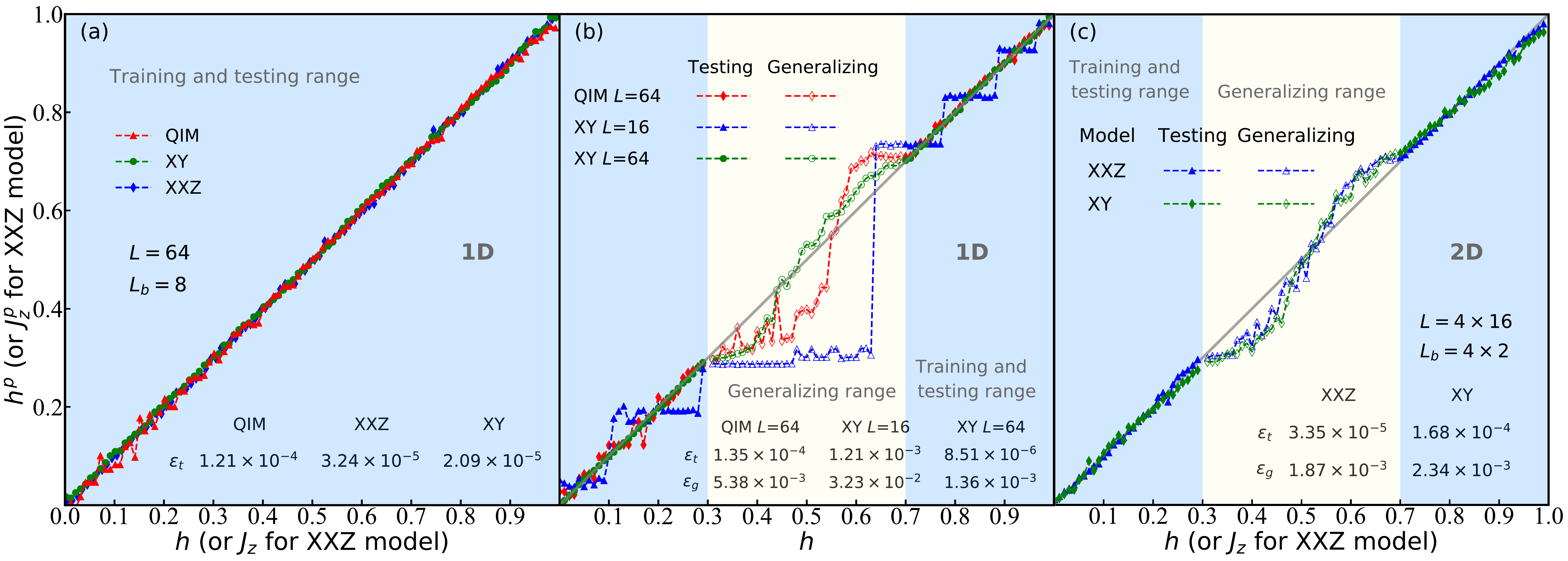}
	\caption{(Color online) The estimated parameters of the Hamiltonian ($h$ or $J_z$) versus their ground truth. The blue shadows indicate the ranges of parameters where we take as the training and testing sets. The yellow shadows indicate the range of the generalizing set. The errors of testing and generalizing sets ($\varepsilon_t$ and $\varepsilon_g$) can be found in the inset tables. (a) Estimations $h^p$ or $J_z^p$ versus the true values $h$ or $J_z$ without generalizing set ($\delta=0$) for the 1D spin models. We take $L=64$ with the subsystem size $L_b=8$. The RDM trick is used. (b) Estimations $h^p$ versus the true $h$ taking $\delta=0.4$ for the 1D spin models. Note the critical point of QIM, $h=0.5$, is put in the middle of the generalizing range. And no RDM trick is used when $L=16$ in the XY model. (c) Estimations $h^p$ versus the true $h$ for the spin models on a $4 \times 16$ square lattice.}
	\label{fig-MainPreds}
\end{figure*}

\section{Results and discussions}

We first benchmark the training and testing accuracies of QubismNet on one-dimensional (1D) QIM by taking $L=64$ as the system size and $L_b=8$ as the subsystem size in the RDM trick. We take the periodic boundary condition, meaning the first and last spins are interacted as nearest neighbors. FIG. \ref{fig-MainPreds} (a) shows the estimated fields $h^p$ against the true fields $h$. The $N_{\text{train}}=1000$ training states are taken as the ground states by uniformly choosing different magnetic fields in $0<h<1$. The $h$'s of the $N_{\text{test}}=100$ testing states are also uniformly taken in $0<h<1$, which are different from the fields of the training states. No generalizing states are taken (i.e., $\delta=0$). The QubismNet accurately estimates the magnetic fields of both the training and testing states. We have the testing error $\epsilon_t \simeq 1.21 \times 10^{-4}$ evaluated by the loss function, i.e., MSE, of all testing states. The estimations are accurate near the critical point, where the states possess relatively long-range correlations.

We also test the QubismNet on 1D XXZ model with periodic boundary condition. We consider two cases, whose Hamiltonians are written, respectively, as
\begin{eqnarray}
	\hat{H}(J_z) = \sum_{\left \langle i, j \right \rangle} (\hat{S}_i^x \hat{S}_j^x + \hat{S}_i^y \hat{S}_j^y + J_z \hat{S}_i^z \hat{S}_j^z), \label{eq-XXZ} \\
     \hat{H}(h) = \sum_{\left \langle i, j \right \rangle} (\hat{S}_i^x \hat{S}_j^x + \hat{S}_i^y \hat{S}_j^y) + h \sum_{k=1}^{L} \hat{S}_k^z. \label{eq-XY}
\end{eqnarray}
The physical parameters to be estimated by the QubismNet are the coupling strength $J_z$ in Eq. (\ref{eq-XXZ}) and the longitudinal field $h$ in Eq. (\ref{eq-XY}). We dub the latter as XY model, where we take zero $J_z$ and non-zero longitudinal field. We use the RDM trick with $L=64$ and $L_b=8$. The testing accuracies of the XXZ and XY models are about $\varepsilon_t \sim O(10^{-5})$. 

\begin{figure}[tbp]
	\centering
	\includegraphics[angle=0,width=1.0\linewidth]{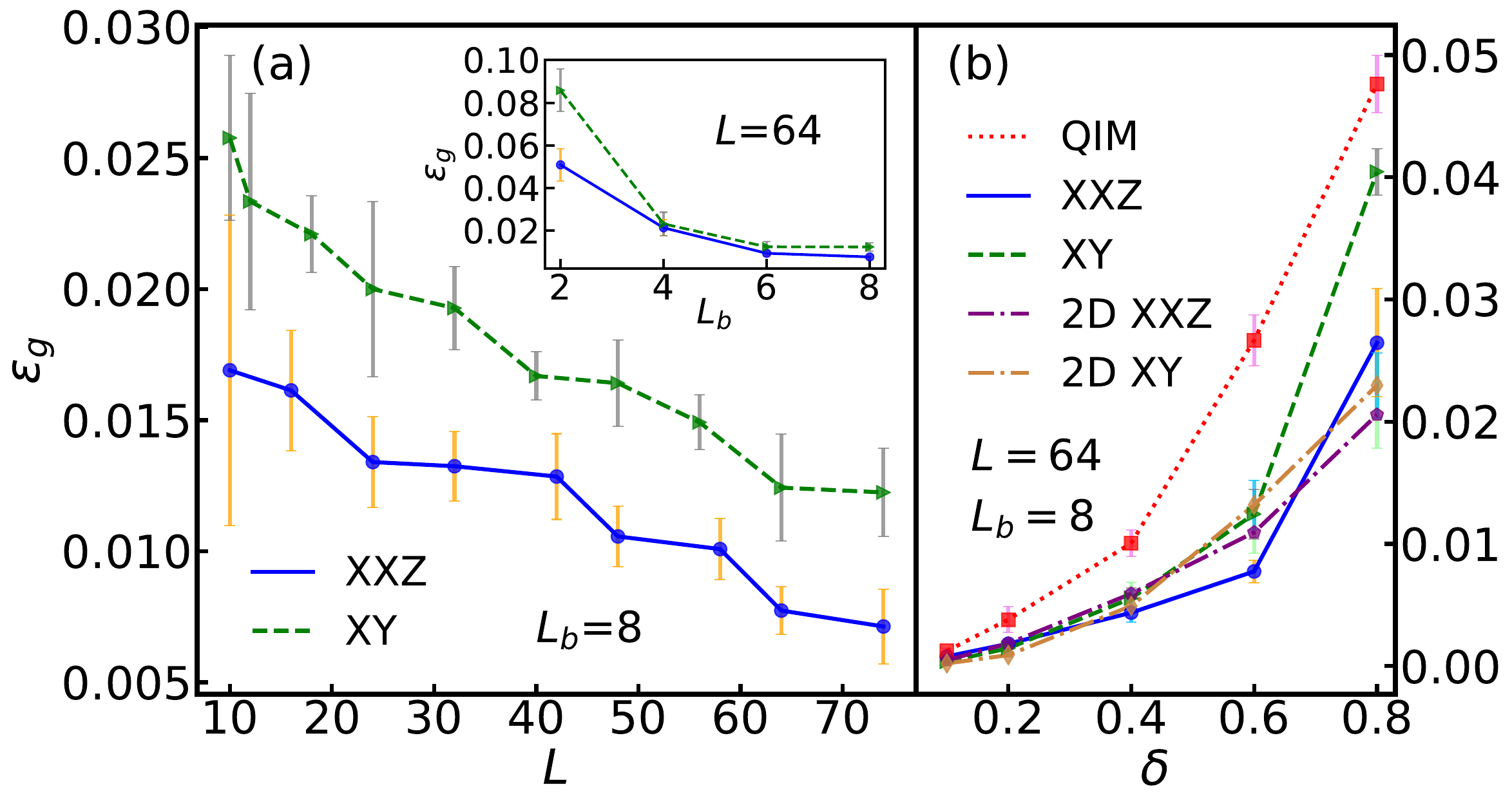}
	\caption{(Color online) (a) The generalizing error $\epsilon_g$ versus the system size $L$ by fixing the generalizing width $\delta=0.6$ and bulk size $L_b=8$ for 1D XXZ and XY models. The inset demonstrates the relation between $\epsilon_g$ and $L_b$ with $L=64$. (b) The relation between $\epsilon_g$ and $\delta$ for 1D and 2D spin models where we fix $L=64$ and $L_b=8$.}
	\label{fig-ErrorBar}
\end{figure}

To benchmark the generalization power, we set $\delta=0.4$ [FIG. \ref{fig-MainPreds} (b)]. Within $0.3<h<0.7$ (the light yellow shadow), no training states are taken. In this range, we averagely take $N_g=40$ $h$'s with an interval $dh=0.01$ as the generalizing set. For the QIM, a quantum phase transition occurs at $h=0.5$. In our setting, the QubismNet only learns from the states away from the critical vicinity. Our results show that it can generalize from what it has learned and estimate the magnetic fields near the critical point. We have the generalizing error (the MSE evaluated by the generalizing set) $\varepsilon_g \sim O(10^{-3})$ using the RDM trick with $L=64$ and $L_b=8$.

For the XY and XXZ models, the system is in the gapless phase for $0<h<1$ \cite{franchini2017introduction}. We set the same ranges for the training, testing and generalizing sets as above. Without the RDM trick, we take $L=16$, and find that ``stages'' appear in the $h$-$h^p$ curves. This is due to the energy gaps caused by the finite-size effects. For instance, by changing $h$ of the XY model from $0.15$ to $0.28$,  no energy level crossing occurs. It means the ground states within this range is the same state.  Such ``stages'' obviously hinder the estimation of the physical parameters from the ground states since QubismNet distinguishes no differences from the states by varying $h$ within a stage. We propose to resolve this issue by increasing $L$ (e.g., to $L=64$). The RDM trick has to be used since we cannot handle the full $2^{64}$ coefficients in the Qubism map. From our results, the ``stages'' of the  $h$-$h^p$ curves are largely suppressed using the RDM trick. The testing and generalizing errors are decreased by more than 100 and 20 times, respectively. QubismNet is also tested on the frustrated breathing kagome antiferromagnet \cite{schaffer2017quantum, repellin2017stability}, where both the testing and generalizing errors are around $O(10^{-4})$. See more details in the Supplementary Material.

We also test on the XXZ and XY models on ($4 \times 16$) 2D square lattice with periodic boundary condition. The subsystem for the RDM is chosen in the middle of lattice with the size $4 \times 2$. The 2D XY model, whose local Hamiltonian is given by Eq. \eqref{eq-XY}, is in an oscillatory phase for $0 \leq h \leq 1$. The XXZ model (Eq. \eqref{eq-XXZ}) is in the paramagnetic phase for $0 \leq J_z \leq 1$. In general, 2D quantum models are much more challenging to simulate. QubismNet works well on such 2D quantum systems as shown in FIG. \ref{fig-MainPreds} (c). With the generalizing width $\delta=0.4$, we get similar performance compared with the chains, with $\varepsilon_t \sim O(10^{-5}) - O(10^{-4})$ and $\varepsilon_g \sim O(10^{-3})$.

To demonstrate the finite-size effects, we show in FIG. \ref{fig-ErrorBar} (a) the $\varepsilon_g$ against $L$ on the XXZ and XY models ($\delta=0.6$). The RDM trick is used with $L_b=8$. The error bars here (and all others in this work) are evaluated by independently and randomly taking the initial values of the variational parameters in the CNN for ten times. The $\varepsilon_g$ of both models are still slightly decreasing with $L$ for $L>60$, meaning it is possible to further reduce the errors by taking larger sizes. In the inset of FIG. \ref{fig-ErrorBar} (a), we fix $L=64$ and find that $\varepsilon_g$ well converges as the subsystem size increases to $L_b>6$ for the models under consideration. Note it would be inefficient to improve the performance by increasing $L_b$ as the complexity will increase exponentially with it. We have tried larger $L_b$ and the results indicate little improvement to the performance.

In FIG. \ref{fig-ErrorBar} (b), we show $\varepsilon_g$ versus $\delta$ for the QIM, XXZ, and XY models. We fix $L=64$ and use the RDM trick with $L_b=8$. Since the QubismNet only learns from the states sampled in $0<h<0.5-\delta/2$ and $0.5+\delta/2<h<1$, it requires more generalization power to estimate the $h$ of the ground states in $0.5-\delta/2<h<0.5+\delta/2$ as $\delta$ increases. Therefore, the generalizing error $\varepsilon_g$ of the QubismNet monotonously increases with $\delta$. But even for $\delta=0.8$, the generalizing error is still insignificant with approximately $\varepsilon_g <0.05$. Meanwhile, the estimations become more fluctuated for larger $\delta$ when randomly initializing the variational parameters of the QubismNet.

\begin{figure}[tbp]
	\centering
	\includegraphics[angle=0, width=1.07\linewidth]{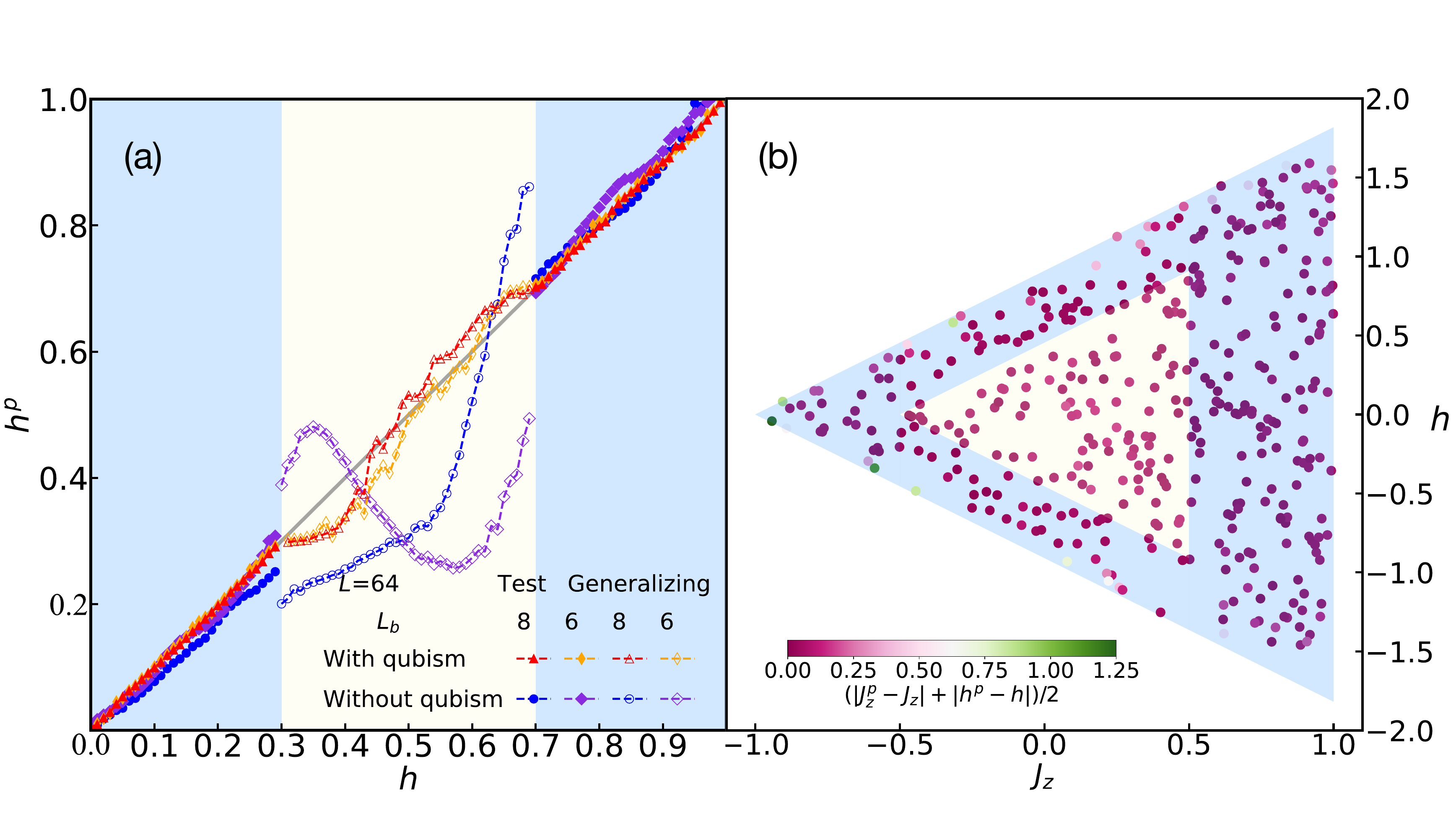}
	\caption{(Color online) (a) The estimations $h^p$ versus the true $h$ on the 1D XY model ($L=64$) with and without the Qubism map. We use the RDM trick with $L_b=8$ and $L_b=6$. (b) The estimations of $h$ and $J_z$ on the 1D XXZ model. The position of each dot shows the ground truth of the two parameters, and its color indicates the error.}
	\label{fig-QubismOrNot}
\end{figure}

In above, we stated that after transforming the states into images, we can take advantage of the power of CNN on processing images. Below, we try to directly reshape the coefficients of a RDM into a $2^{2L_{b}}$-dimensional vector. Then we use a 1D version of CNN, which consists of 1D convolutional and pooling layers, to map the vector to the estimations of the target parameters. FIG. \ref{fig-QubismOrNot} (a) shows the estimations $h^p$ versus $h$ on the 1D XY model with $L=64$ and $L_b=6$ and $8$. The $\varepsilon_t$ and $\varepsilon_g$ without the Qubism map becomes more than ten times larger than that with the Qubism map. These results imply that the Qubism map is a reasonable choice, since the image "visualizes" the physics of the state in the patterns of fractals \cite{rodriguez2012qubism}. We do not exclude the possibilities of other maps that may outperform the Qubism map.

The estimation of two parameters is also tried on 1D Heisenberg XXZ model, where both the magnetic field $h$ and $J_{z}$ which represents the anisotropy are estimated simultaneously. We choose $L=64$ and use the RDM trick with $L_b=8$. As shown in FIG. \ref{fig-QubismOrNot} (b), the training and testing set are obtained within the blue region, and the generalizing set are from the yellow region. The ground state is in a gapless phase. The color of each dots illustrates the error between estimation and label. The generalizing error is $7.54 \times 10^{-3}$.

\section{Summary and perspective}
\label{sec-summary}

Our work is a start-up of using the classical ML models to directly learn the quantum data (e.g., wave-functions or density matrices). CNN models possess high non-linearity, thus it would be interesting to compare with the parameterized quantum circuit models \cite{mitarai2018quantum, liu2018differentiable, zhu2019training} that normally represent unitary transformations on quantum states. Our results suggest the impressive learning and generalization powers of CNN in such issues. It could provide a key tool in designing the Hamiltonian in order to, for instance, prepare target states in Hamiltonian-based quantum simulators \cite{das2008colloquium, georgescu2014quantum}. An important topic for the future investigations is to test the generalization power while breaking the data balance to different extents. Our proposal can be generalized to learn from the experimental data of quantum measurements in, e.g., a quantum state tomography process \cite{vogel1989determination, cramer2010efficient, lanyon2017efficient, xin2019local}. The idea of using NN to inversely solve challenging numeric problems can be potentially generalized, e.g., to the constraint satisfaction problems \cite{cook1971complexity, krzakala2009hiding}.

\section*{Acknowledgments}
SJR is grateful to Ding Liu and Ya-Tai Miu for helpful discussions. SJR is supported by NSFC (Grant No. 12004266 and No. 11834014), Beijing Natural Science Foundation (No. 1192005 and No. Z180013), Foundation of Beijing Education Committees (No. KM202010028013), and the Academy for Multidisciplinary Studies, Capital Normal University. Z. C. Tu is supported by the National Natural Science Foundation of China (Grant No. 11975050).

%


\clearpage

\section*{Supplementary Material}
		
In the supplementary material, more details about the Qubism map and CNN are provided, and the images obtained by the Qubism map from various models in 1D and 2D are shown. The detailed descriptions and some supplementary data on applying QubismNet to 2D quantum systems on the square and breathing kagome lattices are given.

\subsection{Qubism Map}

\begin{figure*}[htbp]
	\centering
	\includegraphics[angle=0,width=0.7\linewidth]{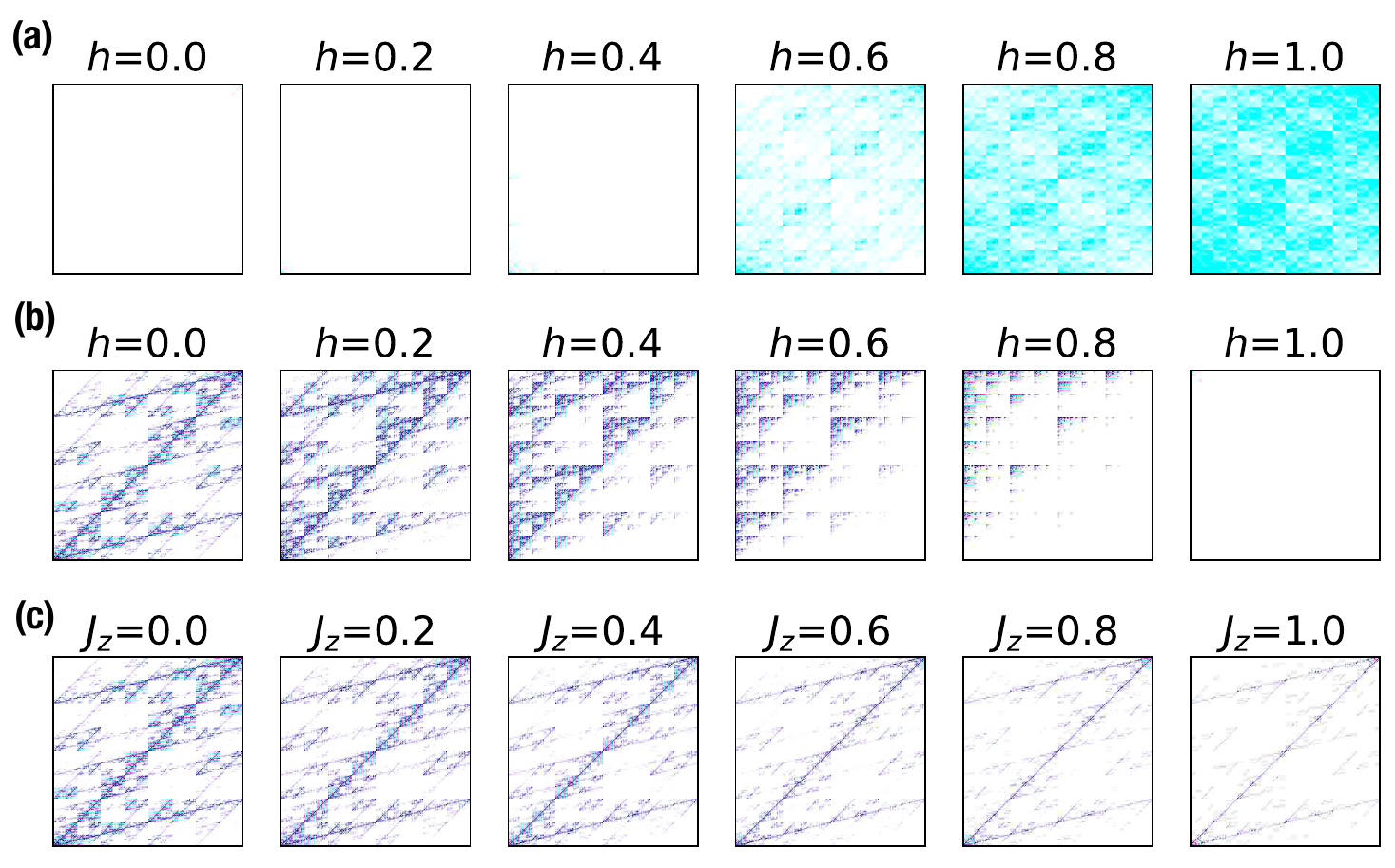}
	\caption{(Color online) Illustrations of the images by the Qubism map from the ground states of the QIM, XY, and XXZ models.}
	\label{fig-exQubism}
\end{figure*}

Consider a quantum system with $L$ spins, denoted by $s=\{ X_1 Y_1 X_2 Y_2 \ldots X_{L/2} Y_{L/2} \}$ with $X_i, Y_i \in \{ 0,1 \}$. The resolution of the image obtained by the Qubism map is $(2^{L/2}, 2^{L/2})$. Each spin configuration corresponds to the pixel in the $x$-th row and $y$-th column of the image satisfying
\begin{equation}
\begin{aligned}
	x &= \sum_{i=1}^{L/2} X_{i} 2^{(L/2-i)}+1, \\
	y &= \sum_{i=1}^{L/2} Y_{i} 2^{(L/2-i)}+1.
\end{aligned}
\end{equation}
The gray-scale value of each pixel is taken as the coefficient of the quantum state in the corresponding spin configuration. Besides, each pixel can be attached to colors based on the phase of quantum state. The images obtained by the Qubism map of 1D QIM, Heisenberg XY, and XXZ models are shown in FIG. \ref{fig-exQubism} (a), (b) and (c), respectively. In all these cases, we take $L= 64$ and $L_b=8$. 

\subsection{Convolutional Neural Network}

\begin{figure*}[htbp]
	\centering
	\includegraphics[angle=0,width=1\linewidth]{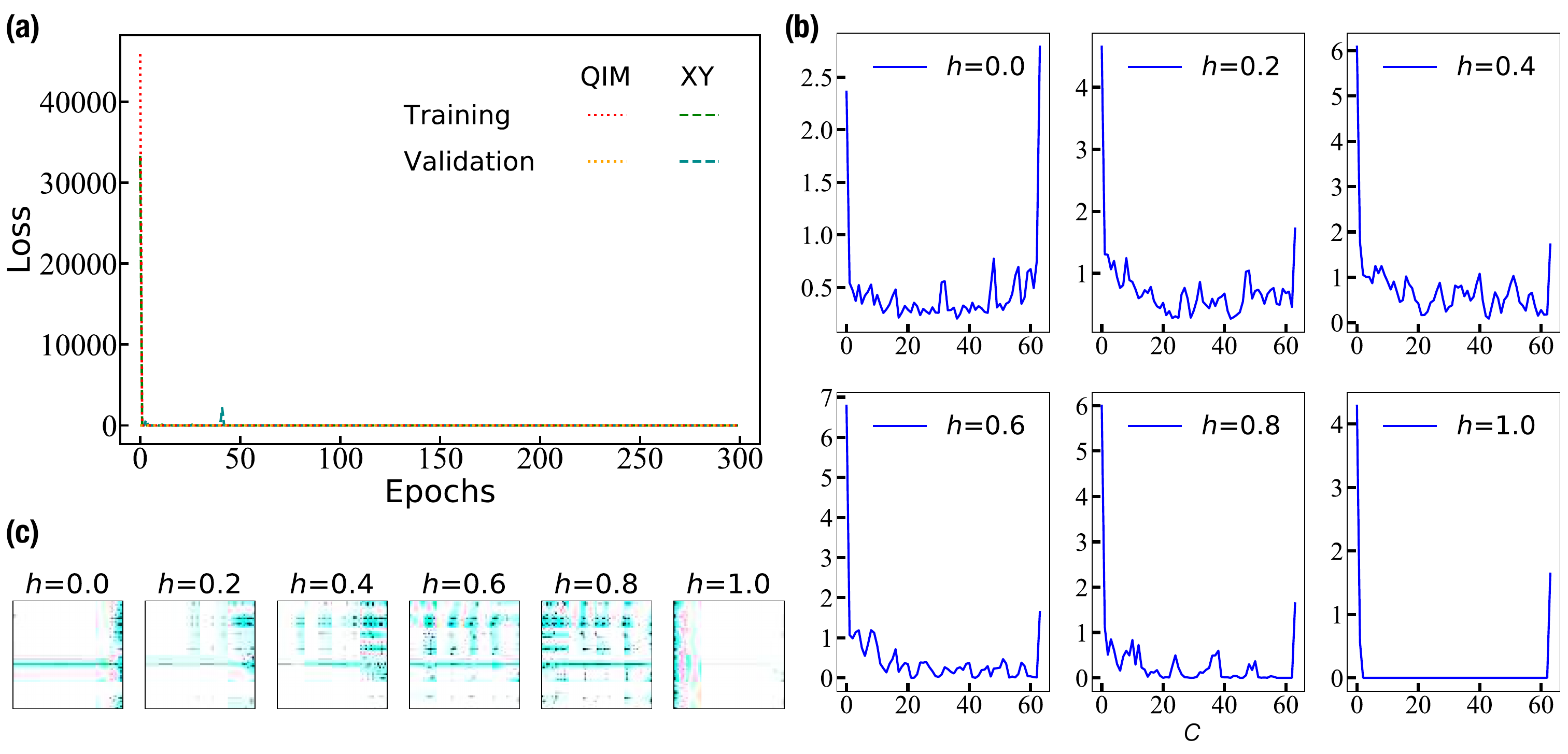}
	\caption{(Color online) The loss functions and extracted features of the CNN. (a) The decreases of the loss function with the training epochs on training and validation set for QIM, XY and XXZ models. (b) For the XY model, we show the average magnitude of the extracted features for different channels $c$. For different $h$, the dominant contribution is always from the first channel. (c) The extracted $64 \times 64$ features in the first channel. In general, we observe that the prominent features (illustrated by the dark dots) move from right to left as $h$ increases.}
	\label{fig-lossVis}
\end{figure*}

CNN has shown the advantage in image recognition and many other challenging tasks. In general, CNN consists of several alternative convolutional and pooling layers, which together serve as a feature extractor. One or many fully-connected layers are then used to map the extracted features to the target output in classification and regression. The CNN we use in this work contains eight layers. The first convolutional layer filters the input images with 32 kernels of size $3 \times 3$ and a stride of size $1 \times 1$. The second convolutional layer take the output of the first convolutional layer as input with same setting as the first convolutional layer. The first max-pooling layer of size $2 \times 2$ follows which downsamples the size of features. Then the third and fourth convolutional layers are used both with 64 kernels of size $3 \times 3$ and a stride of size $1 \times 1$. All convolutional layers use padding around the images so the outputs has the same height/width dimension as the inputs. The second max-pooling layer follows with pool size $2 \times 2$. Next, the output of the second max-pooling layer is flattened and then input to two fully-connected layers with 128 and 32 neurons respectively. The output of the last fully connected layer is fed to one neuron which produce the final output of the physical parameter. Note that we choose two successive convolutional layers to improve the quality of features because of the more complex nonlinearity and larger receptive fields. The rectified linear unit (ReLU) \cite{nair2010rectified} is chosen as activation function for all the convolutional and fully-connected layers. A linear activation function is used for the output layer. The optimizer we use is RMSprop with the initial learning rate 0.001. "He normal initialization" is employed as initialization method which has been proven to get good effect in CNN with the rectifier nonlinearities \cite{he2015delving}.

We split data into four datasets which are training, validation, testing and generalizing set. The training set is used to train the CNN model. The validation set is randomly taken from the training set with the percentage of $10 \%$. The validation set has two roles in our work. One is choosing the proper hyper-parameters such as the number of neurons in each layer, the optimizer, and so on. The other is the early-stopping technique. The testing set and the generalizing set consist of data which never show in the training set. The difference is that data in the testing set have same range of the training set however the generalizing range is beyond the training range.

Two tricks on avoiding overfitting during the training process is employed. One is the dropout \cite{hinton2012improving, srivastava2014dropout} with rate of $p=0.5$, meaning randomly masking $50 \%$ of neurons during training. It is a commonly used method making the NN robust. The other is validation. We always save the best model selected by the minimum loss of the validation set which consists of the samples randomly chosen $10 \%$ from the training set. The best model is then used on the testing and generalizing set. Note that the input images are normalized so that the features are in $[0, 1]$. We construct and train our CNN with Keras \cite{chollet2015keras}, a high-level open-source API running on top of TensorFlow. We train the CNN for 300 epochs for each case (one epoch means training by all the samples in training set once). The total training time is roughly 0.3 hours on a server equipped with NVIDIA P100 graphics processing unit when we use 1000 training samples.

FIG. \ref{fig-lossVis} (a) shows the loss functions versus epochs for the 1D QIM and XY models with $L=64$, $L_b=8$ and $\delta=0.4$. The loss functions of the training and validation sets decrease rapidly and become quite small in a few epochs for all the three models. This indicates that QubismNet is trainable and feasible. Besides, the validation loss functions do not increase during the whole training process implying no overfitting.

Furthermore, we go deep inside the CNN and see extracted features by the convolutional/pooling part of the CNN. After the last pooling layer, each sample is mapped to a $64 \times 64 \times 64$ tensor as the extracted features where 64 is the number of channels. FIG. \ref{fig-lossVis} (b) shows the average magnitude of each channel $c$. We find the the dominant contribution is from the first channel in almost all the cases. In FIG. \ref{fig-lossVis} (c), we demonstrate the $64 \times 64$ features in the first channel. Interestingly, we observe that the prominent features (with larger values marked by the deeper dots) move in general from right to left with the increase of $h$. These results indicate that QubismNet capture some consecutive rule from the quantum states in its own way.

All codes used in the manuscript and this supplementary material can be found on GitHub \cite{QubismNet}.


\subsection{RDM-based method for two-dimensional lattice}

\begin{figure*}[htbp]
	\centering
	\includegraphics[angle=0,width=0.7\linewidth]{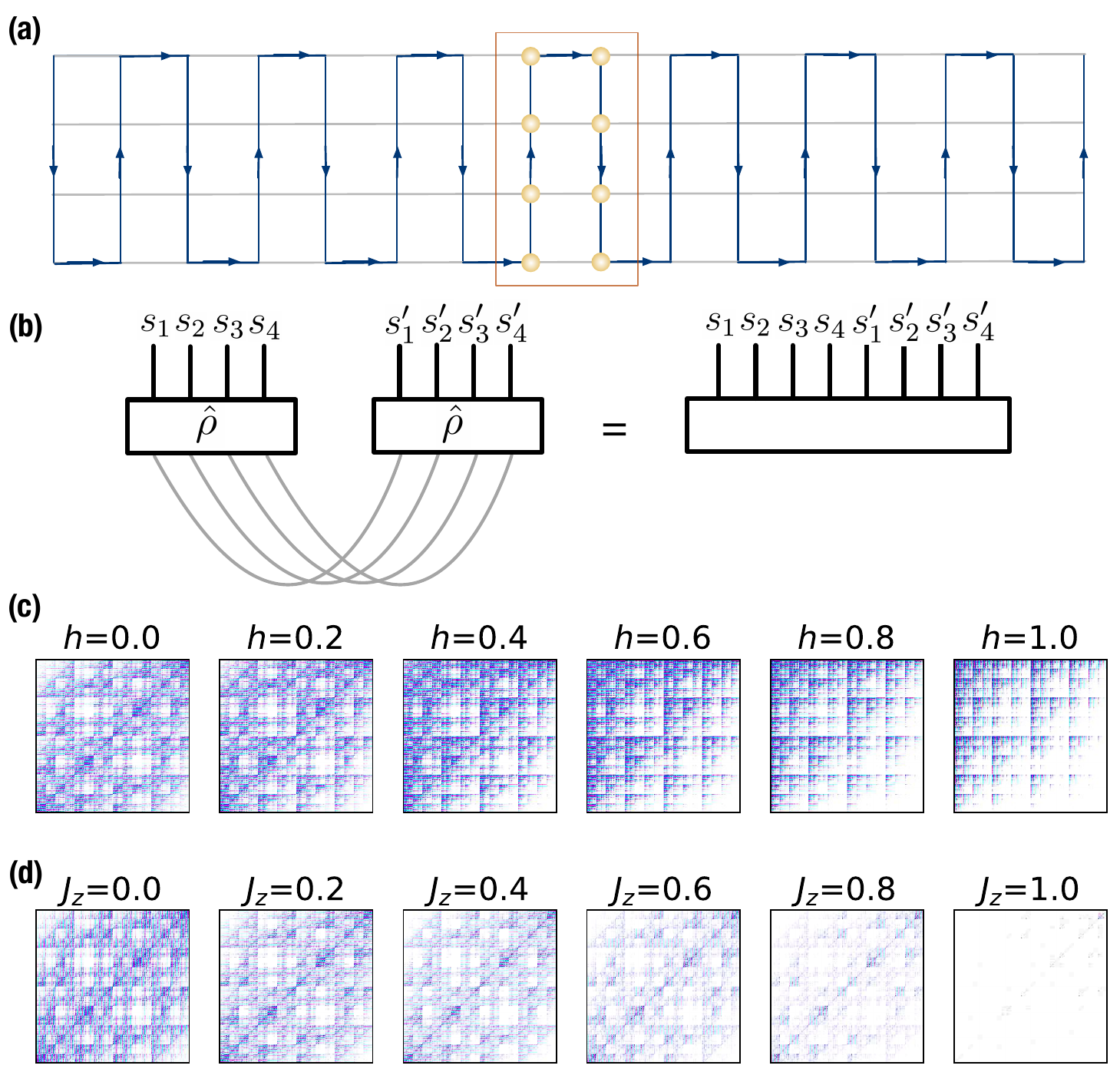}
	\caption{(Color online) An illustration of DMRG applied to the 2D system, and the images obtained from the 2D ground states. (a) An illustration of how the 2D lattice is stretched to a 1D chain in order to use DMRG to simulate the ground states. We choose a $2\times 4$ sub-system in the middle to define the RDM. (b) An illustration of how the reduced density matrices are constructed. (c) The images obtained by applying the Qubism map to the ground states in the 2D XX model in different transverse fields $h$. (d) The images from the 2D XXZ model with different $J_z$.}
	\label{fig-ex2D}
\end{figure*}

We follow the standard recipe of DMRG on solving the ground states of 2D quantum models \cite{SW12DMRG2DRev}. The 2D lattice model with nearest-neighboring interactions is stretched into a chain with long-range interactions, as illustrated in FIG. \ref{fig-ex2D} (a). We set the size of lattice as $4 \times 16$ for both the 2D XY and XXZ models. The subsystem used to calculate RDM is chosen in the middle of lattice with size $4 \times 2$ as illustrated. For the purified state of $\rho^2$, we treat all the degrees of freedom in the \textit{bra} space as one of the two dimensions of the 2D image, and treat those in the \textit{ket} space as the other dimension of the image. This ordering is illustrated in FIG. \ref{fig-ex2D} (b). Several images obtained from the ground states of the 2D XY and XXZ models with different $h$ or $J_z$ are shown in FIG. \ref{fig-ex2D} (c) and (d) as examples. 

\subsection{QubismNet for breathing kagome antiferromagnet}

\begin{figure*}[htbp]
	\centering
	\includegraphics[angle=0,width=1.0\linewidth]{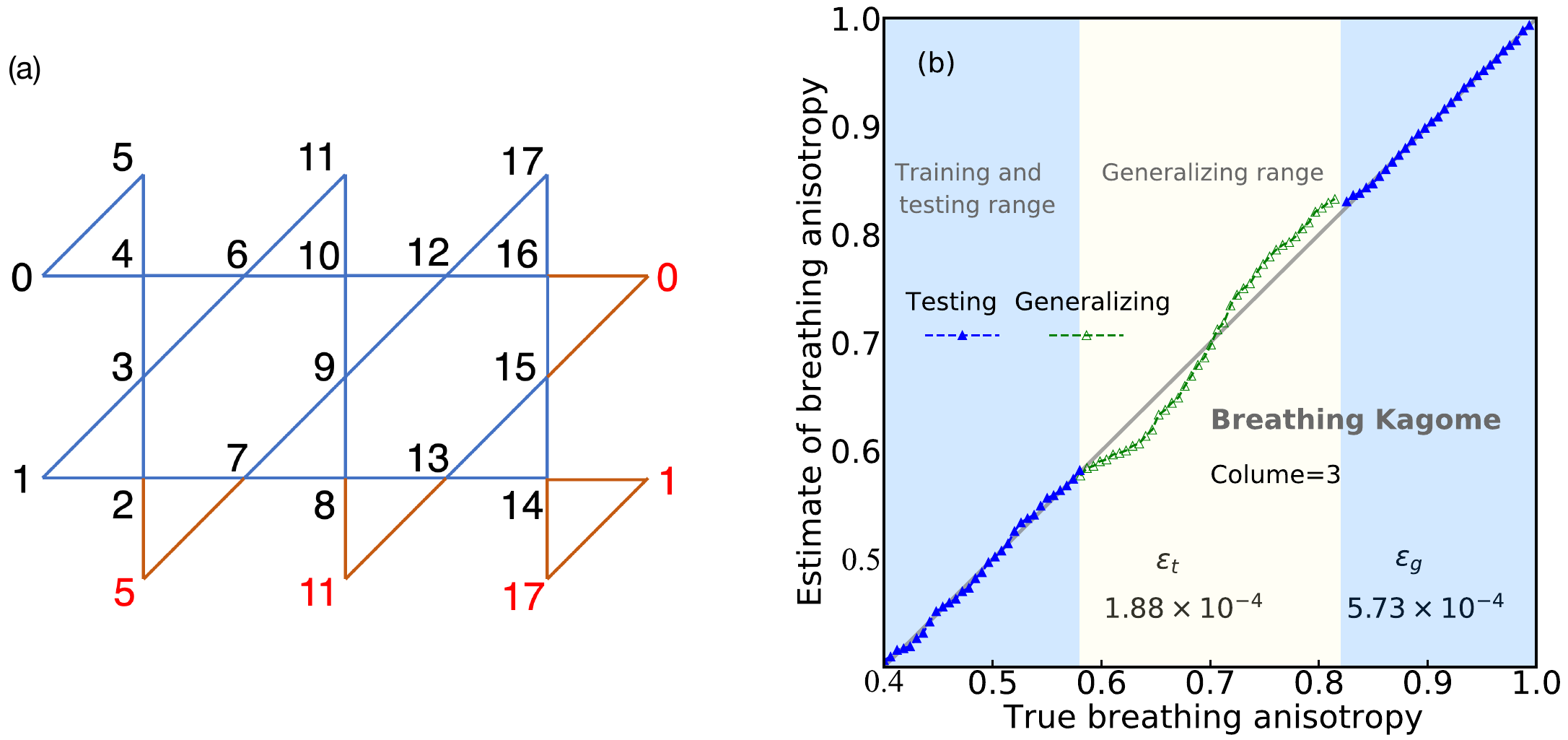}
	\caption{(Color online) (a) An illustration of the breathing kagome antiferromagnet, where the couplings of the upper and lower triangles are antisymmetric. The number of columns is three. Periodic boundary condition is chosen as shown in the red edges. (b) shows the estimations of the breathing anisotropy versus its ground truth. We use the RDM trick with $L_{b} =8$.}
	\label{fig-kagome}
\end{figure*}

We also tried QubismNet for the non-trivial breathing kagome antiferromagnet with the Hamiltonian

\begin{equation}
\begin{aligned}
	\hat{H} = & J_{\triangle} \sum_{\left \langle i, j \right \rangle \in \triangle} (\hat{S}_i^x \hat{S}_j^x + \hat{S}_i^y \hat{S}_j^y + \hat{S}_i^z \hat{S}_j^z)\\
	& +  J_{\nabla} \sum_{\left \langle i, j \right \rangle \in \nabla} (\hat{S}_i^x \hat{S}_j^x + \hat{S}_i^y \hat{S}_j^y + \hat{S}_i^z \hat{S}_j^z)
	\label{eq-BKA}
\end{aligned}
\end{equation}
where $J_{\triangle}$ and $J_{\nabla}$ represent the coupling strength in the up and down triangles, respectively. The illustration of the kagome lattice with periodic boundary condition is shown in FIG. \ref{fig-kagome} (a). The number of columns is three and the image size after the qubism map is $512 \times 512$. The breathing anisotropy $J_{\nabla}/J_{\triangle}$ is chosen to vary between 0.4 and 1.0. The ground states are magnetically disordered RVB states and solved by DMRG algorithm. FIG. \ref{fig-kagome} (b) shows the results of predicting the breathing anisotropy on the testing and generalizing sets. The mean square errors are of the order of magnitude $O(10^{-4})$. This demonstrates the validity of the QubismNet on the nontrivial kagome model with magnetically disordered ground states.

%


\end{document}